\documentclass[12pt]{article}
\usepackage{amssymb}
\usepackage{amsmath}
\usepackage{graphicx}
\usepackage{bm}
\newcommand{\mathsym}[1]{{}}

\usepackage{graphicx}
\usepackage{rotating}
\usepackage{color}
\usepackage{cancel}

\newcommand{\bra}{\begin{array}}
\newcommand{\era}{\end{array}}
\newcommand{\beq}{\begin{equation}}
\newcommand{\eeq}{\end{equation}}
\newcommand{\beqar}{\begin{eqnarray}}
\newcommand{\eeqar}{\end{eqnarray}}

\newcommand{\be}{\begin{equation}}
\newcommand{\ee}{\end{equation}}
\newcommand{\bea}{\begin{eqnarray}}
\newcommand{\eea}{\end{eqnarray}}
\newcommand{\bd}{\begin{displaymath}}
\newcommand{\ed}{\end{displaymath}}

\newcommand{\de }{\Delta }

\newcommand{\lb }{ \left (}
\newcommand{\rb }{ \right )}

\newcommand{\ra }{\mathbb{R}}
\newcommand{\rc }{\mathbb{R}_{\al}}

\newcommand{\p }{ \partial}

\newcommand{\xd }{ \dot{x}}

\newcommand{\op}{ \oplus  }
\newcommand{\om}{ \ominus }
\newcommand{\ot}{ \otimes }
\newcommand{\od}{ \oslash }

\newcommand{\ta }{ \theta}
\newcommand{\ep}{\epsilon}
\newcommand{\al }{\alpha}

\newtheorem{definition}{Definition}[section]

\newtheorem{proposition}{Proposition}[section]

\begin{document}

\vspace{20pt}

\begin{center}

{\LARGE \bf Deformed special relativity based on $\al$-deformed binary operations

\medskip
 }
\vspace{15pt}

{\large  Won Sang Chung${}^{1,\dag}$ and Mahouton Norbert Hounkonnou${}^{2,\dag\dag}$

}

\vspace{15pt}
{\sl ${}^{1}$Department of Physics and Research Institute of Natural Science,\\
 College of Natural Science,\\
 Gyeongsang National University, Jinju 660-701, Korea}\\

{\sl ${}^{2}$ International Chair in Mathematical Physics and Applications,
ICMPA-UNESCO Chair,\\ University of Abomey-Calavi,  072 BP 50, Cotonou, Rep. of Benin}\\

\vspace{5pt}
E-mails:  {$
{}^{\dag}$mimip44@naver.com};
 {$
{}^{\dag\dag}$norbert.hounkonnou@cipma.uac.bj}, with copy to hounkonnou@yahoo.fr
\vspace{10pt}
\end{center}

\begin{abstract}
In this paper,  we  define a new velocity having a dimension of $(Length)^{\al}/(Time)$ and
 a new acceleration having a dimension of $(Length)^{\al}/(Time)^2$, based on the fractional addition rule. We then discuss the fractional mechanics in one dimension. We show  the conservation of fractional energy, and formulate the Hamiltonian formalism for the fractional mechanics. As a matter of illustration, we exhibit some examples for the fractional mechanics.

\end{abstract}

 \today

 \tableofcontents

\section{Introduction}

Recently,  a new analysis, called a pseud analysis, appeared in mathematics [1,2]. It is a generalization of the classical analysis, where instead of the field of real numbers a semiring
is taken on a real interval $[a, b] \subset [-\infty, +\infty]$ endowed with
pseudo-addition $ \op$ and with pseudo-multiplication $\ot,$ with different applications in mathematics and physics, e.g. in modeling nonlinearity, uncertainty in optimization problems, nonlinear partial differential equations, nonlinear difference equations, optimal control, fuzzy systems, decision making , game theory. Its advantage  is that there covered with one theory, as universal mathematical theory, and so with unified methods, problems (usually nonlinear and uncertain) from  various fields. It also gives solutions in the form which are not achieved by other approaches, e. g., Bellman difference equation, Hamilton Jacobi equation with non-smooth Hamiltonians. For more details, see \cite{pap1} and references therein.

\begin{definition}
The pseudo binary operations are defined by the help of  a monotonous bijective map $f,$ called their generator, as:
\bea \label{eq1}
x \op_f y &=& f^{-1} ( f(x) + f(y)) \cr
x \om_f y  &=& f^{-1} ( f(x) - f(y)) \cr
x \ot_f y  &=& f^{-1} ( f(x) f(y) ) \cr
x \od_f y &=& f^{-1}( f(x)/f(y)).\nonumber
\eea
\end{definition}
The simplest choice $f(x)=x$ gives the ordinary binary operations, and
		\begin{eqnarray}
		x^n\oplus_{f}y^m= x^n+ y^m,
		\end{eqnarray}
		\begin{eqnarray*}
		x^n\ominus_{f}y^m= x^n- y^m,
		\end{eqnarray*}
		\begin{eqnarray*}
		x^n\otimes_{f}y^m=x^ny^m
		\end{eqnarray*}
		and 
		\begin{eqnarray*}
		x^n\oslash_{f}y^m=x^n\slash y^m.
		\end{eqnarray*}
Furthermore,
\begin{eqnarray*}
\big(x\oplus_{f}y\big)^n= \sum_{k=0}^{n}{n\choose k}\,x^k\,y^{n-k},
\end{eqnarray*}
\begin{eqnarray*}
\big(x\ominus_{f}y\big)^n= \sum_{k=0}^{n}{n\choose k}\,(-1)^k\,x^k\,y^{n-k},
\end{eqnarray*}
\begin{eqnarray*}
\big(x\otimes_{f}y\big)^n=(xy
)^n,
\end{eqnarray*}
and 
\begin{eqnarray*}
\big(x\oslash_{f}y\big)^n=(x\slash y
)^n.
\end{eqnarray*}

It can be easily checked that the operation $\op_f$ and $ \ot_f $ satisfy the commutativity
and associativity properties. Through the map $f$, we can perform many deformed binary operations [3,4]. The first use of this pseudo binary operations was made by Einstein [5] in the velocity addition. The second use  was made in constructing the $q$-additive entropy theory [6-8].

For
	\begin{eqnarray*}
	f(x)= x^{\alpha},\quad \alpha >0,
	\end{eqnarray*}
	we have:
	\begin{eqnarray*}
	f^{-1}(x)=x^{1\over \alpha}.
	\end{eqnarray*}
	Thus,
	\begin{eqnarray*}
	x^n\oplus_{f}y^m=\big(x^{n\,\alpha}+y^{m\,\alpha}
	\big)^{1\over \alpha}, 
	\end{eqnarray*}
	\begin{eqnarray*}
	x^n\ominus_{f}y^m= \big(x^{n\,\alpha}-y^{m\,\alpha}
	\big)^{1\over \alpha},
	\end{eqnarray*}
	\begin{eqnarray*}
	x^n\otimes_{f}y^m=x^ny^m
	\end{eqnarray*}
	and 
	\begin{eqnarray*}
	x^n\oslash_{f}y^m=x^n\slash y^m.
	\end{eqnarray*}
Moreover,
\begin{eqnarray*}
\big(x\oplus_{f}y\big)^n=\big(x^{\alpha}+y^{\alpha}
\big)^{n\over \alpha},
\end{eqnarray*}
\begin{eqnarray*}
\big(x\ominus_{f}y\big)^n=\big(x-y
\big)^{n\over \alpha},
\end{eqnarray*}
\begin{eqnarray*}
\big(x\otimes_{f}y\big)^n=(x\,y
)^n,
\end{eqnarray*}
and 
\begin{eqnarray*}
\big(x\oslash_{f}y\big)^n=(x\slash y
)^n.
\end{eqnarray*}

Recently, Chung and  Hassanabadi \cite{{ChungHassan}} considered a special choice of $f,$ 

\be \label{eq2}
f(x) =|x|^{\al-1} x , ~~~~ \al>0,
\ee
so that the deformed multiplication and deformed division may be the same as the ordinary ones. 
Using this, these authors
 studied the anomalous diffusion process by using the $\al$-deformed mechanics which possesses the $\al$-translation in space $x \rightarrow x \op \delta x$.
Inserting the equation (\ref{eq2})  into  (\ref{eq1}), we have the $\al$-deformed  binary operations, i. e.  $\al$-addition, $\al$-subtraction, $\al$-multiplication and $\al$-division, as:
\bea
a\op_{\al} b &=& |a |a|^{\al-1}+ b|b|^{\al-1}|^{1/\al-1} ( a |a|^{\al-1}+ b|b|^{\al-1}) \\
a\om_{\al} b &=& |a |a|^{\al-1}- b|b|^{\al-1}|^{1/\al-1} ( a |a|^{\al-1}- b|b|^{\al-1})\\
a\otimes_{\al} b &=& a b\\
a\od_{\al} b &=& \frac{a}{b}.
\eea
Besides, we get the following identities:
\begin{eqnarray*}
a^n\oplus_{\al}b^m=\Big|a^n\,|a^n|^{\alpha-1}+b^m\,|b^m|^{\alpha-1}
\Big|^{1\over \alpha-1}\big(a^n\,|a^n|^{\alpha-1}+b^m\,|b^m|^{\alpha-1}
\big), 
\end{eqnarray*}
\begin{eqnarray*}
a^n\ominus_{\al}b^m=\Big|a^n\,|a^n|^{\alpha-1}-b^m\,|b^m|^{\alpha-1}
\Big|^{1\over \alpha-1}\big(a^n\,|a^n|^{\alpha-1}-b^m\,|b^m|^{\alpha-1}
\big), 
\end{eqnarray*}
\begin{eqnarray*}
a^n\otimes_{\al}b^m=a^nb^m
\end{eqnarray*}
and 
\begin{eqnarray*}
a^n\oslash_{\al}b^m=a^n\slash a^m.
\end{eqnarray*}
Furthermore,
\begin{eqnarray*}
\big(a\oplus_{\al}b\big)^n=\Big|a\,|a|^{\alpha-1}+b\,|b|^{\alpha-1}
\Big|^{n\over \alpha-1}\big(a\,|a|^{\alpha-1}+b\,|b|^{\alpha-1}
\big)^n,
\end{eqnarray*}\begin{eqnarray*}
\big(a\oplus_{\al}b\big)^n=\Big|a\,|a|^{\alpha-1}-b\,|b|^{\alpha-1}
\Big|^{n\over \alpha-1}\big(a\,|a|^{\alpha-1}-b\,|b|^{\alpha-1}
\big)^n,
\end{eqnarray*}
\begin{eqnarray*}
\big(a\otimes_{\al}b\big)^n=(a\,b)^n,
\end{eqnarray*}

\begin{eqnarray*}
\big(a\oslash_{\al}b\big)^n=(a\slash a)^n.
\end{eqnarray*}
Interestingly,  the  multiplication and division are invariant after $\al$-deformation.

The paper is organized as follows. In Section 2, we derive the Newton law of $\alpha-$deformed Newton mechanics. Section 3 is devoted to the characterization of $\al$-deformed Galilean relativity. the $\al$-deformed Galilei group is described,  and energy conservation law is deduced. In Section 4, we study the special relativity with $\al$-translation symmetry. Section 5 deals with an analysis of two body decay.

%
%
%
%
%
%
%
%
%
%
%
%
%
%
%
%
%
%
%

\section{$\alpha-$deformed Newton mechanics}

In an ordinary Newtonian mechanics in one dimension, the Newton velocity is defined as
\be
v = \frac{dx}{dt},
\ee
where $dx$ and $dt$ denote the infinitesimal displacement and infinitesimal time interval, respectively. The infinitesimal displacement is invariant under spacial translation $x \rightarrow x + \delta x$ and the  infinitesimal time interval is invariant under temporal translation $t \rightarrow t + \delta t$.

If we impose new translation symmetry based on $\al$-addition rule, we need  to change the definition of velocity so that it may possess this new symmetry. Here we impose two translation symmetries, i. e. , the  { $\al$-translation in position,} $ x \rightarrow x \op_{\al} \delta x, $
and
 { $\al$-translation in time,} $ t \rightarrow  t \op_{\al} \delta t$.

%
%

In \cite{ChungHassan},
 the authors defined the deformed velocity so that it is invariant under $\al$-translation in position and ordinary translation in time. Then, the average velocity is given by
\be
v_{ave} = \frac{ f_{\al} (x' \om_{\al} x)}{  t' - t}=\frac{\Delta_{\al} x}{ \Delta t} =\frac{|x'|^{\al-1}x' -|x|^{\al-1}x}{t'-t}.
\ee
Taking $t'\rightarrow t$, the velocity is given by
\be
v = \frac{d_{\al} x}{d t} = \al |x|^{\al-1}  \frac{dx}{dt}.
\ee
If we impose the $\al$-translation in both time and position, 
we have to change the definition of the velocity. In this case, the average $\al$-velocity is furnished by the expression
\be
v_{\al,ave} = \frac{ f_{\al} (x' \om_{\al} x)}{  f(t' \om_{\al} t)}=\frac{\Delta_{\al} x}{ \Delta_{\al} t} =\frac{|x'|^{\al-1}x -|x|^{\al-1}x}{|t'|^{\al-1} t' -|t|^{\al-1}t},
\ee
where we call $\Delta_{\al} x$ and $\Delta_{\al} t$ the $\al$-displacement and $\al$-time-interval, respectively.
Taking $t'\rightarrow t$, the $\al$-velocity is yielded by
\be
v_{\al} = \frac{d_{\al} x}{d_{\al} t} = t^{1-\al} |x|^{\al-1}  \frac{dx}{dt}
\ee
Because $v_{\al}$ is $\al$-translation invariant, the $\al$-acceleration is defined as
\be
a_{\al} = \frac{ dv_{\al}}{d_{\al} t} =   \frac{1}{\al} t^{1-\al}   \frac{dv_{\al}}{dt}
\ee
Since the $\al$-velocity and $\al$-acceleration have  dimension $[Length]^{\al}/[Time]^{\al}$ and dimension $[Length]^{\al}/[Time]^{2\al}$, respectively, the Newton equation is obtained by the relation
\be
|F|^{\al-1} F = m^{\al} a_{\al} \label{newton}
\ee
or, equivalently, 
\be
F = m |a_{\al}|^{\frac{1}{\al}-1} a_{\al}
\ee
In  mechanics with $\al$-translation symmetry, the  $\al$-velocity and $\al$-acceleration have the fractional dimensions which are different  from the ordinary ones unless $\al=1$. But, for the force, we assumed that it has the same dimension as the one in the $\al=1-$mechanics.

\section{$\al$-deformed Galilean Relativity }
 Based on the new definition of $\al$-velocity and $\al$-acceleration, we define the $\al$-inertial frames of reference possessing the property that
 a body with zero net force acting upon these frames does not $\al$-accelerate; that is, such a body is at rest or moving at a constant $\al$-velocity. Here we assume
%
{\it the physical laws  must be the same in all $\al$-inertial frames of reference.}


\noindent Now let us consider two inertial frames $S(t,x)$ and $S'(t',x')$ moving at a relative constant $\al$-velocity $u_{\al}$ with $x$-axes. The Newton equation is invariant under the transformations
\be
v_{\al}' = v_{\al} - u_{\al}, ~~~~ v_{\al}'=\frac{d_{\al} x'}{d_{\al} t}
\ee
and
\be \label{relative1}
x' = x \om_{\al} |u_{\al}|^{\frac{1}{\al}-1} u_{\al}t, ~~~~~ t'=t.
\ee

\subsection{$\al$-deformed matrix}

Like the $\al$-deformed binary operations for numbers based on the $\al$-map, we can define the  $\al$-deformed binary operations for matrices  based on the $\al$-map.
Now let us consider $( m \times n)$ matrix
 \be
  {A} =\begin{pmatrix}a_{11}&a_{12}&\cdots &a_{1n}\\a_{21}&a_{22}&\cdots &a_{2n}\\\vdots &\vdots &\ddots &\vdots \\a_{m1}&a_{m2}&\cdots &a_{mn}\end{pmatrix}=\left(a_{ij}\right)
 \ee
 We define the  $\al$-map of the matrix ${A}$ as
 \bd
   f(A) :=\begin{pmatrix}f(a_{11})&f(a_{12})&\cdots &f(a_{1n})\\f(a_{21})&f(a_{22})&\cdots &f(a_{2n})\\\vdots &\vdots &\ddots &\vdots \\f(a_{m1})&f(a_{m2})&\cdots &f(a_{mn})\end{pmatrix}
   \ed
   \be
   = \begin{pmatrix}|a_{11}|^{\al-1}a_{11}&|a_{12}|^{\al-1}a_{12}&\cdots &|a_{1n}|^{\al-1}a_{1n}\\|a_{21}|^{\al-1}a_{21}&|a_{22}|^{\al-1}a_{22}&\cdots &|a_{2n}|^{\al-1}a_{2n}\\\vdots &\vdots &\ddots &\vdots \\|a_{m1}|^{\al-1}a_{m1}&|a_{m2}|^{\al-1}a_{m2}&\cdots &|a_{mn}|^{\al-1}a_{mn}\end{pmatrix}
 \ee
 and its inverse as
 \bd
   f^{-1}(A) :=\begin{pmatrix}f^{-1}(a_{11})&f^{-1}(a_{12})&\cdots &f^{-1}(a_{1n})\\f^{-1}(a_{21})&f^{-1}(a_{22})&\cdots &f^{-1}(a_{2n})\\\vdots &\vdots &\ddots &\vdots \\f^{-1}(a_{m1})&f^{-1}(a_{m2})&\cdots &f^{-1}(a_{mn})\end{pmatrix}
   \ed
   \be
   = \begin{pmatrix}|a_{11}|^{1/\al-1}a_{11}&|a_{12}|^{1/\al-1}a_{12}&\cdots &|a_{1n}|^{1/\al-1}a_{1n}\\|a_{21}|^{1/\al-1}a_{21}&|a_{22}|^{1/\al-1}a_{22}&\cdots &|a_{2n}|^{1/\al-1}a_{2n}\\\vdots &\vdots &\ddots &\vdots \\|a_{m1}|^{1/\al-1}a_{m1}&|a_{m2}|^{1/\al-1}a_{m2}&\cdots &|a_{mn}|^{1/\al-1}a_{mn}\end{pmatrix}
 \ee
  The scalar multiplication of the matrix is the same as the one in $\al=1-$theory. We define  the $\al$-addition and  $\al$-subtraction of two matrices of same type $A, B$  as
 \be
 A \op_{\al} B: = f^{-1} ( f (A) + f(B))
 \ee
 \be
 A \om_{\al} B := f^{-1} ( f (A) - f(B)),
 \ee
 which imply
 \be
 (A \op_{\al} B )_{ij} = a_{ij} \op_{\al} b_{ij}
  \ee
 \be
 (A \om_{\al} B )_{ij} = a_{ij} \om_{\al} b_{ij}
 \ee
 For $n \times p$ matrix $A$ and $ p \times m$ matrix $B$, the $\al$-multiplication of $A$ and $B$ is defined as
  \be
 A \ot_{\al} B := f^{-1} ( f(A) f(B))
 \ee
 which implies
 \be
 (A \ot_{\al} B)_{ij}  = \bigoplus_{k=1}^p a_{ik} b_{kj}
  \ee
 where
 \be
 \bigoplus_{k=1}^p  C_k = C_1 \op_{\al} C_2 \op_{\al} \cdots \op_{\al} C_p
 \ee

 \subsection{ $\al$-deformed Galilei group}

 Based on the $\al$-operations for matrices, we can rewrite the eq.(\ref{relative1}) as
 \be
 \begin{pmatrix} x'\\ t'\end{pmatrix} = T_{\al} ( u_{\al}) \ot_{\al} \begin{pmatrix} x\\ t\end{pmatrix}
 = \begin{pmatrix} 1 & - |u_{\al}|^{\frac{1}{\al}-1} u_{\al} \\ 0 & 1\end{pmatrix} \ot_{\al} \begin{pmatrix} x\\ t\end{pmatrix}
 \ee
 Here we know that  the transformation matrix $T_{\al} ( u_{\al})$ forms a Lie group with the $\al$-multiplication. Indeed,   the following properties are satisfied:
\begin{itemize}
\item $T_{\al} ( u_{\al}) \ot_{\al} T_{\al} ( v_{\al})= T_{\al} ( u_{\al} + v_{\al}) $.
\item  The $\al$-multiplication is associative.
\item The identity is $T_{\al} (0)$.
\item The inverse is $T_{\al} (-u_{\al})$.
\end{itemize}

\subsection{ Energy conservation}

Because $dx$ is not invariant under the $\al$-translation, we use  $\al$-translational invariant infinitesimal displacement as $d_{\al} x = \al |x|^{\al-1} dx$ in defining the work,
\be\label{work}
|W|^{\al-1} W = - \int d_{\al} x |F|^{\al-1} F,
\ee
having the same dimension asin the $\al=1-$mechanics.
 We define the potential energy through  the conservative force,
\be
|F|^{\al-1} F = - \frac{d_{\al}U}{d_{\al} x} = - |x|^{1-\al}|U|^{\al-1}  \frac{d U}{dx}.
\ee
Thus, for the conservative force,  we have
\be
|W_{1\rightarrow 2}|^{\al-1} W_{1\rightarrow 2} = - ( |U_2|^{\al-1} U_2  - |U_1|^{\al-1} U_1).
\ee
Inserting the Newton equation (\ref{newton}) into (\ref{work}), we get
\be
|W_{1\rightarrow 2}|^{\al-1} W_{1\rightarrow 2} = K_2 - K_1
\ee
where the kinetic energy is given by
\be
K = \frac{1}{2} m^{\al} v^2_{\al}.
\ee
Considering the dimension, the conservation of energy is provided by
\be
|E|^{\al-1} E = K + |U|^{\al-1} U  =  \frac{1}{2} m^{\al} v^2_{\al}  + |U|^{\al-1} U  =\frac{ p_{\al}^2}{2m^{\al}} + |U|^{\al-1} U,
\ee
where the linear momentum is expressed as
\be
p_{\al} = m^{\al} v_{\al}.
\ee
The energy has the same dimension as  in the $\al=1-$mechanics, while the linear momentum has fractional dimension.

\section{ Special relativity with $\al$-translation symmetry}

The 3-position in non-relativistic mechanics is changed into  4-position (or event) in the relativistic one. Let us consider the event  $P ( c t, x, y, z)$, where $c$ is  the Newton speed of light, (i. e. speed with $\al=1$). Based on the definition of  $\al$-translation invariant  infinitesimal displacement and $\al$-translation invariant  infinitesimal time interval, the
$\al$-translation invariant  distance ($\al$-distance) of infinitesimally close space-time events denoted by $ds_{\al}$  is
given by
\be
d_{\al}s^2 = c^{2\al}  d_{\al}t^2 -  d_{\al}x^2 - d_{\al}y^2 -  d_{\al}z^2.
\ee
The $\al$-deformed  proper-time $\tau_{\al}$ is
\be
d_{\al}\tau^2  = \frac{d_{\al}s^2}{c^{2\al}}.
\ee

\subsection{$\al$-Lorentz transformations }

The $\al$-Lorentz transformations  making  invariant the space-time interval 
\be
( \Delta_{\al} s)^2
= \lb {c^{\al}}(  |t|^{\al-1} t )\rb^2
- \lb  ( |x|^{\al-1} x )\rb^2
\ee
are given by
 \bd
 |x|^{\al-1} x = c^{\al} |t'|^{\al-1} t' sh_{\al} (\psi) + |x'|^{\al-1} x' ch_{\al} (\psi)
 \ed
 \be
 c^{\al}|t|^{\al-1} t =  c^{\al}|t'|^{\al-1} t' ch_{\al} (\psi) + |x'|^{\al-1} x' sh_{\al} (\psi),
 \ee
 where the little $\al$-deformed hyperbolic functions are defined by
 \be
 sh_{\al} (\psi) : =\frac{1}{2} \lb e_{\al} (\psi ) - e_{\al} (-\psi )\rb
 = \sinh \lb  |\psi|^{\al-1} \psi \rb
 \ee
 \be
 ch_{\al} (\psi) : =\frac{1}{2} \lb e_{\al} (\psi ) + e_{\al} (-\psi )\rb =  \cosh \lb |\psi|^{\al-1} \psi \rb
 \ee
 \be
 th_{\al} (\psi) := \frac{sh_{\al} (\psi) }{ch_{\al} (\psi) } =  \tanh \lb  |\psi|^{\al-1} \psi \rb
 \ee
 \be
e_{\al} (x) := e^{ |x|^{\al-1} x}.
\ee
 The  little $\al$-deformed hyperbolic functions obey the relations
 \be
 ch_{\al}^2 (\psi)-sh_{\al}^2 (\psi)=1.
 \ee
 In terms of the $\al$-deformed binary operations, we get
 \bd
 x = c t' Sh_{\al} (\psi ) \op x' Ch_{\al} (\psi )
 \ed
 \be
 c t = c t' Ch_{\al} (\psi ) \op x' Sh_{\al} (\psi ),
 \ee
 where the big $\al$-deformed hyperbolic functions are
 \be
 Ch_{\al} (\psi) := \left| ch_{\al} (\psi)\right|^{\frac{1}{\al} -1} ch_{\al} (\psi)
 \ee
 \be
 Sh_{\al} (\psi) := \left| sh_{\al} (\psi)\right|^{\frac{1}{\al} -1} sh_{\al} (\psi)
 \ee
 \be
  Th_{\al} (\psi) := \frac{Sh_{\al} (\psi) }{Ch_{\al} (\psi) }.
 \ee
obeying
\be
|Ch_{\al} (\psi)|^{2} \om |Sh_{\al} (\psi)|^{2}=1
\ee
Consider in the coordinate system $(ct, x)$ the origin of the coordinate system $(ct', x')$. Then, $x'=0,$ and
\bd \label{coordinate1}
x = c t' Sh_{\al} (\psi )
 \ed
 \be
 c t = c t' Ch_{\al} (\psi ).
 \ee
 Dividing the two equations gives
 \be
 \frac{x}{ c t} =Th_{\al} (\psi ),
 \ee
 or,
  \be
 \frac{|x|^{\al-1}x}{ c^{\al} |t|^{\al-1} t} =th_{\al} (\psi ).
 \ee
 Since $ \frac{ |x|^{\al-1}x}{ |t|^{\al-1} t} =v_{\al}$ is the relative uniform $\al$-velocity (see \cite{chung1}) of the two systems, we  identify the physical meaning
of the imaginary "rotation angle $\psi$" as
\be
th_{\al} (\psi ) = \frac{v_{\al}}{ c^{\al}} =\beta_{\al}.
\ee
Using the following identities
\be
ch_{\al} (\psi ) = \gamma_{\al}, ~~
sh_{\al} (\psi) = \gamma_{\al} \beta_{\al},,
\ee
where
\be
\gamma_{\al} = \frac{1}{\sqrt{ 1 - \beta_{\al}^2}},
\ee
we obtain the $\al$-deformed Lorentz transformation of the form
\bd
 |x|^{\al-1} x = \gamma_{\al} \lb |x'|^{\al-1} x'  + v_{\al} |t'|^{\al-1} t' \rb
 \ed
 \be
 |t|^{\al-1} t = \gamma_{\al} \lb |t'|^{\al-1} t'  + \frac{v_{\al} }{c^{2\al}} |x'|^{\al-1} x' \rb.
 \ee
 Expressing the eq.(\ref{coordinate1}) in terms of the $\al$-deformed binary operations, we get
 \bd
 x = \Gamma_{\al} \lb  x' \op v_{\al}^{1/\al} t' \rb
 \ed
 \be
  t = \Gamma_{\al} \lb  t' \op \frac{v_{\al}^{1/\al}}{c^{2}}  x' \rb
 \ee
 where
 \be
 \Gamma_{\al} = \gamma_{\al}^{1/\al} = ( 1 - \beta_{\al})^{-\frac{1}{2\al}}.
 \ee
 If we set
 \be\label{speed1}
 u_{\al} = \lb \frac{|x|^{\al-1}}{|t|^{\al-1}} \rb \frac{dx}{dt}, ~~~
  u'_{\al} = \lb \frac{|x'|^{\al-1}}{|t'|^{\al-1} } \rb \frac{dx'}{dt'}
 \ee
the addition of $\al$-velocity becomes
\be
u_{\al} = \frac{ u_{\al}' + v_{\al}}{ 1 + \frac{ v_{\al} u_{\al}}{c^{2\al}}}
\ee
If we regard the $\al$-speed of light as $c^{\al}$, the eq.(\ref{speed1}) shows that the $\al$-speed of light remains invariant, and, hence, the speed of light also remains invariant under the $\al$-deformed Lorentz transformation.

\subsection{$\al$-Lorentz group}


Now, let us introduce the four $\al$-velocity. For that, we change the notation as:
\be
ct = x^0, ~~x=x^1, ~~y=x^2, ~~ z=x^3
\ee
Then,  the four $\al$-velocity is given by
\be
u_{\al}^a = \frac{|x^a|^{\al-1} dx^a}{ (\tilde{d} \tau)_{\al}}
\ee
or, explicitly,
\be
u^0_{\al} = c^{\al} \gamma_{\al}
\ee
\be
u^i_{\al} = v_{\al}^i  \gamma_{\al},\;\; i= 1, 2, 3.
\ee
Therefore,  we have
\be
\eta_{ab} u_{\al}^a u_{\al}^b = c^{2\al}.
\ee

\subsection{ Energy and $\al$-momentum}

The four $\al$-momentum is defined as
\be
p_{\al}^a = m^{\al} u^a_{\al}
\ee
explicitly giving
\be
p^0_{\al} = m^{\al} c^{\al} \gamma_{\al}
\ee
\be
p^i_{\al} = m^{\al} v_{\al}^i  \gamma_{\al}.
\ee
Thus, we have
\be\label{energy2}
\eta_{ab} p_{\al}^a p_{\al}^b =m^{2\al} c^{2\al}.
\ee
Here, we have $p_{\al}^a \ne ( E/c, \vec{p}_{\al})$ because the energy in $\al$-deformed mechanics  has the same unit as  in the undeformed case. Therefore, we set
\be
p_{\al}^a =  \lb \lb \frac{E}{c}\rb^{\al}, \vec{p}_{\al}\rb
\ee
Thus, the eq.(\ref{energy2}) gives
\be
E^{2\al} = c^{2\al} |\vec{p}_{\al}|^2 + m^{2\al} c^{4\al}
\ee
When $|\vec{v}_{\al}|\ll c^{\al}$, we have
\be
E^{\al} \approx \frac{|\vec{p}_{\al}|^2}{ 2 m^{\al}}
\ee
which is the same as the non-relativistic case.


\section{Two body decay }

The simplest kind of particle reaction is the two-body decay of unstable particles. A well
known example from nuclear physics is the alpha decay of heavy nuclei. In particle physics, one
observes, for instance, decays of charged pions or kaons into muons and neutrinos, or decays of
neutral kaons into pairs of pions, etc. 

Consider the decay of a particle of mass $M$ which is initially at rest. Then, its four $\al$-momentum
is $P  = (M^{\al}, \vec{0})$, where we set $c=1$. This reference frame is called the centre-of mass frame (CMS). Denote the four $\al$-momenta of the two daughter particles by $p_1 = (E_1^{\al}, \vec{p}_{\al,1}), p_2 = (E_2^{\al}, \vec{p}_{\al,2})$. From the momentum conservation,  we get
\be
\vec{p}_{\al,1} + \vec{p}_{\al,2}=0
\ee
The energy conservation is 
\be
M^{\al} = \sqrt{ |\vec{p}_{\al,1}|^2 + m_1^{2\al}} + \sqrt{ |\vec{p}_{\al,2}|^2 + m_2^{2\al}}
\ee
If we set
\be
p = |\vec{p}_{\al,1}| = |\vec{p}_{\al,2}|, 
\ee
we have 
\be
p = \frac{1}{2M^{\al}} \sqrt{ ( M^{2\al} - ( m_1^{\al} - m_2^{\al})^2 )( 
M^{2\al} - ( m_1^{\al} + m_2^{\al})^2 )}
\ee
Thus, we have 
\be
M \ge  m_1 \op_{\al} m_2. 
\ee

\end{document}